# Topics in Present-day Science Technology and Innovation: Ultrafast Relaxation Processes in Semiconductors

*Clóves Gonçalves Rodrigues[a]\*, Áurea Rosas Vasconcellos[b], Roberto Luzzi[b]*

[a]*Departamento de Física, Pontifícia Universidade Católica de Goiás – PUC Goiás, CP 86, CEP 74605-010, Goiânia, GO, Brazil*
[b]*Departamento de Física da Matéria Condensada, Instituto de Física "Gleb Wataghin", Universidade de Campinas - Unicamp, CEP 13083-859, Campinas, SP, Brazil*

The nowadays notable development of all the modern technology, fundamental for the progress and well being of world society, imposes a great deal of stress in the realm of basic Physics, more precisely on Thermo-Mechanical Statistics. In electronics and optoelectronics we face situations involving physical-chemical systems far-removed-from equilibrium, where ultrafast (in pico- and femto-second scale) and non-linear processes are present. Here we describe in an extended overview the question of ultrafast relaxation processes in the excited plasma in semiconductors.

## 1. Introduction

Processes involved in nowadays advanced technology are mostly associated to optical and transport properties in systems far-away-from equilibrium, displaying ultrafast (pico to femto-second time scales) relaxation processes, being in constrained geometries (nanometric scales), and presenting nonlinear dynamic behavior. This is to be expected in multiple situations involving technological applications and their end use in manufacturing processes.

This sets a certain stress on the associated basic Physics, or more specifically on nonequilibrium statistical thermo-mechanics, and we may mention the topics of[1]

a) Ultrafast Relaxation Processes in Semiconductors
b) Nonlinear Transport in Highly-Polar Semiconductors
c) Low-Dimensional Complex Semiconductors
d) Nonlinear Higher-Order Thermo-Hydrodynamics
e) Nonequilibrium Bose-Einstein-like Condensation and Complexity in Condensed Matter
f) Thermo-Statistics of Complex Structured Systems
g) Nonconventional Thermo-Hydrodynamics

In this Feature Article we consider in detail the thermo-statistical aspects of the ultrafast evolution of the nonequilibrium state of highly photoexcited polar semiconductors under high levels of excitation.

These processes can be evidenced, and its evolution followed, in experiments of ultrafast laser spectroscopy. On this we reproduce parts of the Introduction of the article "Big Payoffs in a Flash" by J. M. Hopkins and W. Sibbett in Scientific American, September 2000 issue, pages 54 to 61, namely

> "How long did it take you to read this sentence? Just recognizing the first letter took only milliseconds. Around 0.05 millisecond, or 50 microseconds, passes each time chemicals diffuse across a synapse, carrying a signal from one neuron to another in your brain. Are you holding the magazine at a comfortable reading distance? It takes light one or two nanoseconds to travel from the page to your eye and about 20 picoseconds to pass through the lens in your eye. And yet these brief natural events are epically long compared with the shortest man-made events, which proceed 1,000-fold more swiftly: pulses of laser light that last for on1y a few femtoseconds (quadrillionths of a second). The science and technology of ultrashort-pulse lasers have enjoyed much exciting progress since they were developed in the mid-1960s. In particular, the past decade has seen pulses shorter than 10 femtoseconds and the emergence of a new generation of versatile, compact ultrashort-pulse lasers – a revolutionary change from their large, temperamental, power-hungry ancestors. Such laser designs, which make use of sophisticated nonlinear optical phenomena and concurrent advances in diode lasers, increasingly meet the stringent specifications and reliability necessary for many industrial and medical applications. As we enter the 21st century, ultrashort-pulse lasers are becoming more impressive in scope and intensity, producing beams that span the electromagnetic spectrum from X-rays to T-rays (terahertz radiation, beyond infrared) and generating optical peak powers as colossal as petawatts (billions of megawatts). As a result, many new applications in physics, chemistry, biology, medicine, and digital optical technology are emerging and attracting worldwide interest in science and industry."[2]

Studies of the optical and transport properties of semiconductors under high levels of excitation have shown a pleiad of novel and quite interesting features evidenced in ultrafast laser spectroscopy (UFLS). This powerful experimental technique involves the interaction of matter with radiation, which is one of the most studied areas of physics and has played a decisive role in the development of modern physics. However, until the second half of the 20th century, all processes investigated have been associated with weak radiation fields for which the usual perturbation theory, and the accompanying linear response theory near equilibrium, is applicable. Although this approach has been remarkably successful in explaining a great variety of phenomena, in the last several decades the new and greatly improved technological situation involving the advent of lasers, providing us with sources of intense electromagnetic radiation, requires new and sophisticated theoretical approaches, that is, a response theory capable to deal with arbitrarily far-from-equilibrium systems. Moreover, the notable improvements in time-resolved laser spectroscopy have made it a very useful tool to be used with a high degree of confidence in the investigation of very rapid microscopic mechanisms in the physical and biological realms[2-6].

\*e-mail: cloves@pucgoias.edu.br

In particular, ultrafast responses and functioning under far-from-equilibrium conditions in semiconductor systems pose new, interesting, and quite engaging problems in the physics of condensed matter. These systems, as we have already emphasized in several occasions, become an extremely useful testing ground for theoretical ideas in the domain of nonequilibrium statistical thermodynamics of many-body systems. Besides the interest in the comprehension of the basic physical principles underlying these significant situations, there exists a parallel relevant technological interest arising out of the fact that semiconductors working in nonequilibrium conditions have multiple practical applications in electronic devices. Picosecond and femtosecond laser spectroscopy allows to probe ultrafast nonlinear irreversible processes in matter, thus providing an extremely adequate and sophisticated experimental instrument for the study of the nonequilibrium thermodynamic evolution of highly excited semiconductor samples[7-13].

The theories appropriate for the treatment of these far-from-equilibrium many-body systems need make it possible to determine the detailed time evolution of the nonlinear irreversible processes that take place in the system while it is probed. This is a quite attractive and actual problem connected with the nonequilibrium nonlinear statistical mechanics and thermodynamics of dynamical processes. UFLS studies of the highly photoexcited plasma in semiconductors (HEPS, which consists of electron and hole pairs – as mobile carriers – created by an intense laser pulse which are moving in the background of lattice vibrations) have received particular attention since the 1970's. These studies provide information on the kinetic of relaxation of the photoexcited carriers and of the nonequilibrium phonon field, as well as on ultrafast transient transport.

Before proceeding further, and closing this Introduction, we recall our previous statement that we are dealing with open systems in far-from-equilibrium conditions, thus requiring a theoretical approach based on a nonlinear nonequilibrium thermodynamics, and an accompanying statistical mechanics, able to account for the rapid time evolution of the dissipative processes that unfold in these systems. In this context it is quite appropriate to quote Ryogo Kubo's statement in the opening address of the Oji Seminar[14]: "statistical mechanics of nonlinear nonequilibrium phenomena is just in its infancy [and] further progress can only be hoped by close cooperation with experiment". UFLS is a precious tool to carry on Kubo's advice, and we may add that since 1978, the year the above sentence was expressed, a good deal of progress have been attained in the development of nonequilibrium statistical mechanics and irreversible thermodynamics. Concerning the first one we may mention the construction of a generalized Gibbs-Boltzmann- style formalism which can be encompassed within Jaynes' Predictive Statistical Mechanics[15,16,17], the so-called Non-Equilibrium Statistical Ensemble Formalism (NESEF for short), reviewed in Luzzi et al.[18,19]. The NESEF is a powerful formalism that provides an elegant, practical, and physically clear picture for describing irreversible processes, adequate to deal with a large class of experimental situations, as an example in semiconductors far from equilibrium, obtaining good agreement in comparisons with other theoretical and experimental results[20-35]. On the other hand, irreversible thermodynamics is dealt with at the phenomenological level in the framework of several approaches, a quite promising one consisting in the so-called Extended Irreversible Thermodynamics[36], and its microscopic foundation at the mechano-statistical level, namely the so-called Informational Statistical Thermodynamics (sometimes referred to as Information-Theoretic Thermodynamics)[37-43].

## 2. Highly Excited Plasma in Semiconductors (HEPS)

Let us explicitly consider the case of highly excited plasma in semiconductors. These are quite interesting physical systems, among other reasons because of the flexibility in the choice of a number of parameters such as Fermi energy, plasma frequency, energy dispersion relations, cyclotron frequency, different types of carriers and effective masses, etc.[44]. In this plasma in solid state the presence of the lattice introduces noticeable differences in comparison with a gaseous plasma. Typically it produces a background dielectric constant of the order of 10, and exciton effective masses one tenth the value of the free electron mass. Thus, the characteristic atomic units for length and energy, the Bohr radius and the Rydberg, become the excitonic radius, $r_X$, and excitonic Rydberg, $R_Y^X$, which are roughly 100 times larger and a hundreth times smaller than the corresponding atomic units respectively. Hence, the so called metallic densities, namely, situations when the intercarrier spacing, $r_s$, measured in units of $r_X$, is in the range 1 to 5[44-47], arise at quite accessible laboratory conditions for concentrations of roughly $10^{16}$ carriers per $cm^3$ and up.

In Figure 1 we depict the situation to be expected in a typical pump-probe experiment. It describes a sample consisting of a direct-gap polar semiconductor where a concentration *n* of electron-hole pairs is generated by a pulse of intense laser light. Direct absorption of one photon occurs if $\hbar\omega_L > E_G$ where $\omega_L$ is the laser frequency and $E_G$ the semiconductor energy gap. Excitation by means of nonlinear effects such as two-photon absorption or second harmonic generation, ($\hbar\omega_L < E_G$ but $2\hbar\omega_L > E_G$) allows for bulk excitation with a good degree of homogeneity. The sample is illuminated by a second laser (probe) of weak intensity, so as to avoid any noticeable modification of the nonequilibrium state of the system produced by the intense pulse from the pumping laser, and an optical response is recorded. Measurements of luminescence do not require a laser probe.

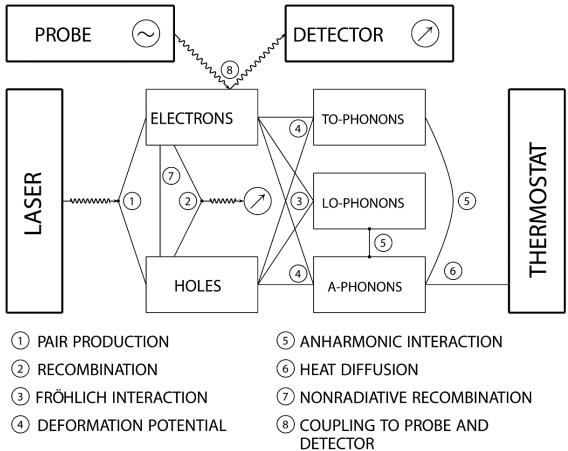

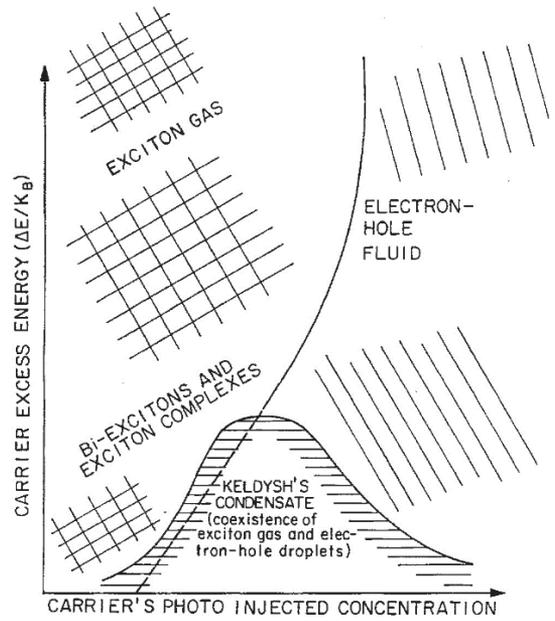

**Figure 1.** Schematic description of a pump-probe experiment in UFLS performed on a semiconductor sample. Several channels involving energy transfer are indicated.

On absorption of the pumped laser light the electrons make transitions from the valence band to the conduction band. These carriers (electrons and holes), with a concentration $n$, are initially narrowly distributed around the energy levels centered on, say, $\epsilon_e$ in the conduction band and $\epsilon_h$ in the valence band, with $\epsilon_e - \epsilon_h \simeq \hbar\omega_L$ (or $2\hbar\omega_L$). Next they are rapidly redistributed in energy space due to the strong long-range Coulomb interaction among them[48,49,50]. Figure 2 provides a description of the nonequilibrium phase diagram of a photoexcited semiconductor.

Once the experiment is set, as described by Figure 1, the connection with theory proceeds via a response function theory. For systems slightly deviated from equilibrium exact close expressions for their response functions to mechanical perturbations can be obtained in the form of correlation functions in equilibrium[51]. A practical way to calculate them is the double-time thermodynamic Green function formalism of Bogoliubov and Tyablikov, described in an already classic paper by Zubarev[52,53]. The actual calculation may be difficult for the case of interacting many-body systems but it is formally closed at this level. However, measurements can be performed on systems strongly departed from equilibrium, when the responses of the systems depend on their instantaneous and local nonequilibrium state, as it is the case of highly excited semiconductors when spectra obtained by means of ultrafast laser spectroscopy depend on the characteristics of the nonequilibrium distributions of the elementary excitations during the lapse of instrumental-resolution time.

Once a HEPS has been created, to describe its macroscopic state it is a prerequisite to define the basis set of macrovariables, that allow to characterized the nonequilibrium thermodynamic state of the system, relevant to the application of the NESEF formalism. As a general rule, according to the formalism one chooses a set of variables to which one has a direct or indirect access in the measurement procedure restricted by the experimental set up. In Figure 1 the main energy relaxation channels between the sample subsystems an between these and external reservoirs (thermostat and pumping laser) are indicated. Since the recorded spectra is averaged over the finite volume observed by the measuring apparatus there is no experimental access to the local hydrodynamic properties of the HEPS; this would be the case when performing space-resolved measurements. The characterization of the macroscopic state of the system is a crucial step in the theory. This question has been discussed elsewhere[18,19,54,55], and, for the particular case of the HEPS in Vasconcellos et al.[56]. The Figure 3 reproduces the chain of successive contracted descriptions appropriate for HEPS.

**Figure 2.** Schematic description of the nonequilibrium phase diagram of the carrier system in a photoexcited semiconductor. In the ordinate, the effective temperature amounts to a measure of the kinetic energy in excess of equilibrium pumped by the laser source.

We have then to consider five steps, namely:

### 2.1. The initial stage

At this point the state of the system is highly correlated, the nonequilibrium processes are determined by very many independent quantities, and there is no satisfactory way to deal with the problem in this stage.

### 2.2. First kinetic stage

In HEPS we need to consider several nonequilibrium subsystems, composing the electron fluid and the ions. As well known, the dynamics of the latter can be described in terms of normal modes of vibration, or, in quantized form, by the different branches of acoustic and optical phonons. There remain the relatively weak anharmonic and electron-phonon interactions. Hence, practically, from the very outset the phonon system can be described in terms of one-particle reduced density functions. A different situation arises for the electron fluid as a result of the presence of the strong Coulomb interaction. In this case, we must first look for a characteristic time, say, $\tau_1 \sim r_0/v_{av}$, to elapse for a description in terms of a one-particle reduced density function to be possible. In this expression, $v_{av}$ is the average velocity which can be estimated to be the Fermi velocity $v_F = (\hbar/m^*)(3\pi n)^{1/3}$ in the highly degenerate limit or the thermal velocity $v_{th} = \sqrt{3k_B T/m^*}$ in the classical limit. For the collision length $r_0$, we can think of the Fermi-Thomas screening length, $\lambda_{TF} = \sqrt{\varepsilon_0 m^* v_{th}^2 / 6\pi n e^2}$, in the first case and Debye-Hückel screening length, $\lambda_{DH} = \sqrt{\varepsilon_0 m^* v_{th}^2 / 4\pi n e^2}$, in the second case[57]. In these expressions $e$ is the absolute value of the electron charge, $m^*$ the carrier (electron or hole) effective mass, $\varepsilon_0$ is the static dielectric background constant and $n$ the concentration of photoinjected carriers. It follows that in both extreme limits one finds that $\tau_1 \sim 1/\omega_{pl}$, where $\omega_{pl} = \sqrt{4\pi n e^2/\varepsilon_0 m^*}$ is the plasma frequency. This is a quite interesting result implying that we may expect to describe the electron fluid in terms of one-particle reduced density matrix in time scales of the order of the period of an electronic plasma wave. In other words, the first kinetic stage for the electron system sets in on time scales necessary for the onset of collective motion. For typical semiconductors $\varepsilon_0 \sim 10$, $m^* \sim 0.1$, and then:

$$\tau_1 \sim \frac{1}{\omega_{pl}} \sim \frac{1.8 \times 10^{-5}}{\sqrt{n}} \quad (1)$$

where $n$ is the number of carriers per cubic centimeter. For several values of the concentration we find the following values: (1) $n = 10^{17}$cm$^{-3}$, $\tau_1 \sim 6 \times 10^{-14}$s; (2) $n = 10^{18}$cm$^{-3}$, $\tau_1 \sim 1 \times 10^{-14}$s; (3) $n = 10^{19}$cm$^{-3}$, $\tau_1 \sim 6 \times 10^{-15}$s, and for very high concentrations, say $n = 10^{22}$cm$^{-3}$, it follows that $\tau_1 \sim 1 \times 10^{-16}$s. This tells us that with laser spectroscopy, attaining time resolutions in the femtosecond time scale, the usual single-particle description of the electron system in semiconductors (solids in general) may become inappropriate for not sufficiently high densities.

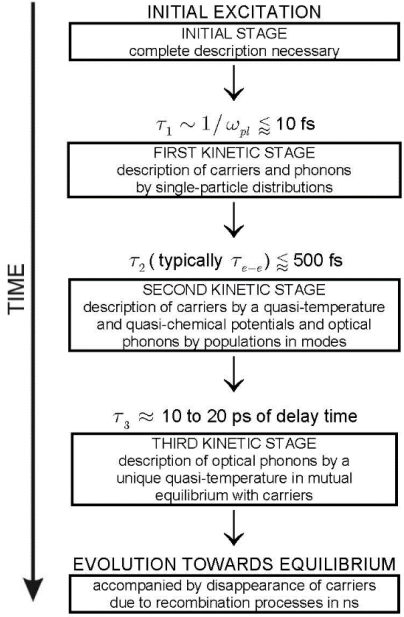

**Figure 3.** Schematic description of the successive kinetic stages in HEPS along its evolution in a pump-probe experiment.

When the single-particle description can be used, the electrons are described in terms of Landau quasi-particles forming a Fermi fluid[58] in single electron Bloch-band states. It can be notice that the production of photoinjected carriers in HEPS modifies the single-electron Bloch energy bands. In particular, a shrinkage of the forbidden energy gap should occur, which can be put into evidence in measurements of the luminescence spectra[59].

Finally, it should be remarked that Landau's Fermi liquid approach to the electron system (quasi-particle and collective excitations) has been extremely successful in dealing with the physical behavior of solid state matter in normal, magnetic, and superconducting states.

### 2.3. Second kinetic stage

Because of the strong Coulomb interaction we may expect a very short time between intercarrier collisions mediated by it, and then a further contraction in the description of the macrostate of the carriers' subsystem should be possible. For the case of homogeneous HEPS, one may expect that such a description can be made in terms of the diagonal elements of the one-particle reduced density function, i.e. the number occupation function $f(\epsilon;t)$, where $\epsilon$ is the energy of the quasi-particle state. This nonequilibrium distribution can be fully characterized by the nonequilibrium thermodynamic variables that NESEF introduces, and whose evolution while the experiment is being performed – evolution which is a consequence of the dissipative effects that are developing in the system – , is completely determined within the scope of the nonlinear quantum kinetic theory based on NESEF as previously described. It is worth recalling that the Lagrange multiplier associated with the energy operator can be interpreted as the reciprocal of a nonequilibrium thermodynamic intensive variable playing the role of a temperature-like quantity, usually referred to as a *quasitemperature*. Hence, within NESEF is given a more rigorous meaning to such concept which was introduced on phenomenological basis, and used in different contexts, by several authors: H. B. G. Casimir and F. K. du Pre[60] for nuclear spins; C. S. Wang-Chang and G. E. Uhlenbeck[61] for molecules; L. D. Landau[62] for plasmas; H. Fröhlich[63] for electrons excited in strong electric fields; V. A. Shklovskii[64] for electrons in superconductors; J. Shah and R. C. C. Leite[65] for photoexcited carriers; J. Shah, R. C. C. Leite, and J. F. Scott[66] for photoexcited phonons. We also call the attention to the fact that HEPS are, at the microscopic quantum-mechanical level, dealt with in the Random Phase Approximation (RPA, sometimes referred to as the generalized time-dependent Hartree-Fock approach); on this we can highlight the books by D. Pines and P. Noziéres[67], Hedin & Lindquist[47], among others. Hence, the description of the system can be done in terms of only the single-particle reduced density function. For the carriers, which are fermions, it follows for $f(\epsilon;t)$ that

$$f_{e(h)}(\epsilon;t) = \frac{1}{1+\exp\{\beta^*(t)(\epsilon - \mu^*_{e(h)}(t))\}} \quad (2)$$

which resembles a time-dependent Fermi-Dirac distribution, with $\beta^*(t)$, and $\mu^*(t)$ connected to the average energy and density by

$$E(t) = \int_0^\infty \epsilon[g_e(\epsilon)f_e(\epsilon;t) + g_h(\epsilon)f_h(\epsilon;t)]d\epsilon \quad (3)$$

$$n(t) = \frac{1}{V}\int_0^\infty g_e(\epsilon)f_e(\epsilon;t)d\epsilon = \frac{1}{V}\int_0^\infty g_h(\epsilon)f_h(\epsilon;t)d\epsilon \quad (4)$$

Given the amount of pumped energy and carrier concentration at a given time, Equations 3 and 4 allow for the determination of the corresponding intensive thermodynamic variables (Lagrange multipliers) $\beta^*(t)$, $\mu^*_e(t)$ and $\mu^*_h(t)$. $V$ is the volume of the system, and $g(\epsilon)$ is the density of states function. The time $\tau_2$, or carrier-carrier collision time, for the establishment of this second kinetic stage can be estimated theoretically and experimentally under given conditions. Measurements of gain spectra in platelets of CdS, like the one given in Figure 4 show a long tail in the low-energy side of the spectrum, below the position (indicated by an arrow) that should correspond to the forbidden energy gap[68]. It arises as a result of the nonstationary character of the carrier energy states, due to relaxation resulting from interactions with impurities and phonons, and the Coulomb interaction between them. The first two interactions are expected to be small (undoped sample and low lattice temperature) compared with the last one. Thus, a fitting of the observed spectrum (full continuous line) assigning a width $\Gamma$ to the one-particle energy levels permits to estimate $\tau_2 \sim \hbar/\Gamma \sim 4\times 10^{-13}$s, i.e. roughly a half picosecond.

Elci et al.[48] have also evaluated, for the case of GaAs, the carrier-carrier collision time using the Golden Rule averaged over the nonequilibrium macrostate using different forms of distributions of carriers in the energy space, in the usual conditions of pump-probe experiments. They obtained values for $\tau_2$ which are also of the order of a fraction of picosecond.

Finally, Collet et al.[49] have calculated the time evolution of $f_e(\epsilon;t)$ in GaAs, starting from an initial condition with electrons and holes distributed in a small region in energy around the values in energy of the conduction and valence states separated by photon energy $\hbar\omega_L$. According to their results, the carrier distribution function evolves to achieve a near Fermi-Dirac-like distribution in roughly $5 \times 10^{-13}$s for $n = 3\times 10$cm$^{-3}$, and roughly $10^{-12}$s for concentrations in the range of $10^{16}$cm$^{-3}$.

All these results show then that in typical conditions, say $n \gtrsim 10^{16}$cm$^{-3}$ and moderate to high-excitation energies, internal thermalization of the carrier subsystem in HEPS follows in the subpicosecond time scale, thus allowing for a contracted description of its macroscopic state in terms of the quasitemperature and the two quasi-chemical potentials.

Moreover, using this description for the electron subsystem it is possible to obtain a very good agreement between calculation and experimental data in experiments of luminescence in CdS[59], where a single-particle approximation was used together with Hedin's proposal[69]. Figure 5 shows recorded spectra and the theoretical curves; the carrier quasi-temperature $T^*$ is obtained from the slope of the high-energy side of the spectrum. The successive shoulders on the low-energy side of the spectrum are longitudinal-optical (LO) phonon-assisted replicas; the finite lifetime of the elementary excitations, not considered in the calculations, should smooth out these shoulders bringing them in better accord with the experimental data. Arrows indicate the position of the peak in the luminescence spectra, whose shift is shown in Figure 6, where $\kappa$ is the concentration dependent Fermi wave number multiplied by the electron spacing $r_S$ in units of the exciton Bohr's radius. Hedin's approach[69] was used to calculate the forbidden energy gap shrinkage due to Coulomb correlations, and this together with the additional shift produced by the superposition of the phonon-assisted replicas produces a very good agreement with experiment.

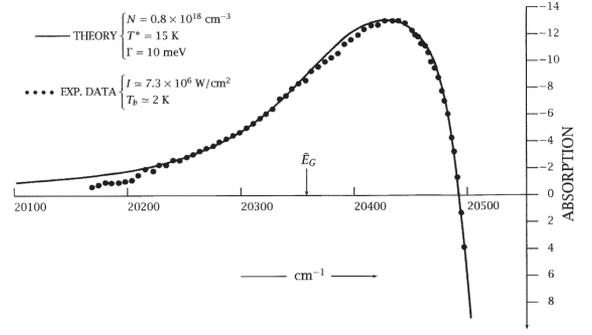

**Figure 4.** Gain-absorption spectrum of CdS in pump-probe experiment. Values of the laser power $I$, reservoir temperature $T_B$, carrier concentration $n$, carrier quasitemperature $T^*$, and reciprocal of the carrier lifetime, $\Gamma$, are shown in the upper left inset. The position of the renormalized band gap is indicated. (Motisuke[68]).

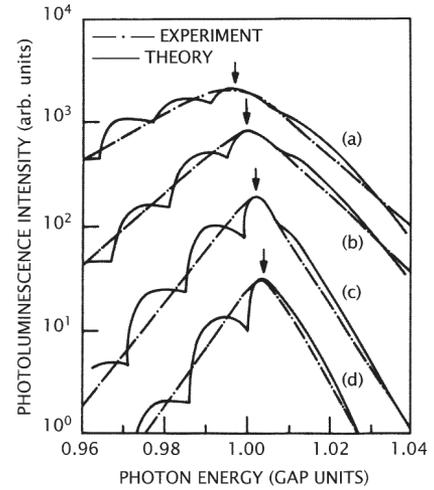

**Figure 5.** The theoretical and experimental luminescence spectra of CdS for increasing, from (d) to (a), laser-pumping power: (a) $\kappa = 0.96$, $T^* = 1790$ K, (b) $\kappa = 0.64$, $T^* = 680$K, (c) $\kappa = 0.48$, $T^* = 295$K, (d) $\kappa = 0.32$, $T^* = 170$ K. Parameter $\kappa$ is the Fermi wave number multiplied by $r_s$, the electron spacing in units of the excitonic Bohr radius. The arrow indicates the position of the peak value (Meneses[59]).

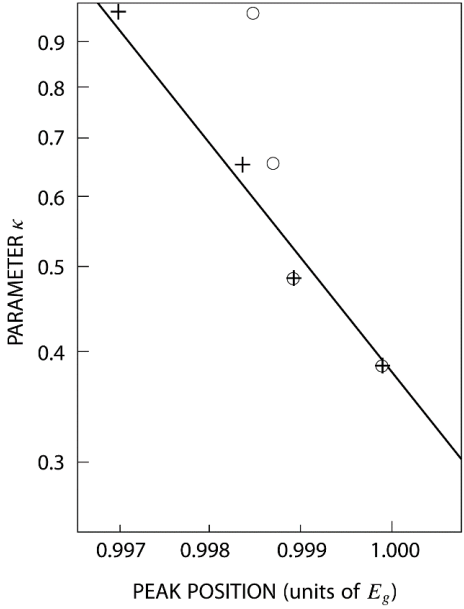

**Figure 6.** In the case of the experimental results, it is shown here the calculated peak position of the luminescence spectra of CdS as a function of the parameter κ: (o) including Coulomb correlation only, and (+) idem plus electron-phonon as described in the text. Parameter κ is the Fermi wavenumber multiplied by $r_S$, the electron spacing in units of the excitonic Bohr radius.

Concerning the phonon subsystem, anharmonic interaction between phonons is weak and so we cannot expect a very rapid thermalization of them. Furthermore, the different branches have different behavior, and the interaction with carriers plays an important role. Interaction of carriers with optical phonons (via deformation potential and Fröhlich interaction) produces a rate of energy transfer much higher than that due to the interaction of carriers with acoustic phonons[70]. Hence, the transfer of the excess energy of the carriers to the thermal bath follows through the indirect channel carriers-to-optical phonons, from the latter to acoustic phonons (via anharmonic interactions) and finally to the thermal reservoir (via heat diffusion).

Figure 7 show the evolution of the LO phonon population (expressed in terms of a quasitemperature $T_q^*(t)$ for a few values of the phonon wavevector in the case of GaAs, as described in Vasconcellos & Luzzi[71]. For these same three modes, it can be shown that the rate of energy transfer from the "hot" carriers to the LO phonons shows a rapid energy transfer in the first few picoseconds, followed by a slower one when a near thermalization of carriers and LO phonons occurs in, roughly, twenty to thirty picoseconds as can be concluded from Figure 7. Comparison with experimental results is hindered by the fact that the latter are scarce as a result of the difficulties to perform Raman scattering experiments (which provide the values of $v(\mathbf{q})$ with a time resolution in the subpicosecond time scale).

We can then conclude that in this second kinetic stage there occurs an internal thermalization of the carrier system, but during the application of the laser pulse, and an interval thereafter, no contraction of description of the optical phonon subsystem can be introduced, and its macroscopic state should be described by the whole set of population distribution functions $v_\eta(\mathbf{q})$.

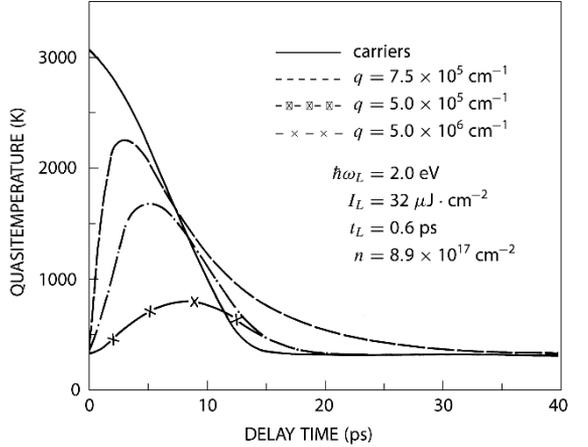

**Figure 7.** Evolution of the quasitemperature of the carriers and three LO-phonon modes, for the conditions indicated in the upper right inset.

### 2.4. Third kinetic stage

Inspection of Figure 7 tells us that near-internal-thermalization of all the optical modes occurs in, roughly, the tens of picoseconds scale after the end of the laser pulse. The internal thermalization of the optical phonon system is then a result of the mutual thermalization of all the optical phonon modes with the carrier subsystem. At this point, a further contraction of the optical phonon system can be introduced, with its macroscopic state characterized by unique quasitemperatures $T_{LO}^*(t)$ and $T_{TO}^*(t)$, with the phonon populations given by the distributions

$$v_{LO(TO)}(\mathbf{q}) = \frac{1}{e^{\hbar \omega_{LO(TO)}(\mathbf{q})/k_B T_{LO(TO)}^*(t)} - 1} \quad (5)$$

attaining a form resembling a time-dependent Planck distribution.

We then have a macroscopic description of the nonequilibrium HEPS in terms of the six macrovariables $T_c^*(t)$, $T_{LO}^*(t)$, $T_{TO}^*(t)$, $T_{AC}^*(t)$, $\mu_e^*(t)$ and $\mu_h^*(t)$. (As already noted, in the usual experimental conditions a very small departure from equilibrium of the acoustic AC phonons is expected). Figure 8 shows the evolution of the carrier quasi-temperature under the conditions of the experiment of Seymour et al.[72], calculated solving the equations for $v_q$ coupled to that of $T_c^*(t)$, $\mu_e^*(t)$ and $\mu_h^*(t)$ (Figure 8b), and the results of another calculation using unique optical-phonon quasi-temperatures $T_{LO}^*(t)$ and $T_{TO}^*(t)$ (Figure 8a) from the outset[70]. Figure 8b agrees quite well with the experimental data, as shown later on in Figure 9. Figure 8 clearly shows the large discrepancy between the curves obtained using the complete contraction of the phonon system, i.e. when using the third stage in place of the second. This is a result of the fact that the use of unique quasi-temperatures for LO and TO phonons overestimates the energy transfer from the carrier system. The selective transfer of energy to the phonon modes (small at low and high wavevector modes and large in a restricted off-center region of the Brillouin zone) produces a much slower cooling down of the carriers. Thus, the third kinetic stage in HEPS, characterized by the internal thermalization of optical phonons, follows at a time $\tau_3$ of, roughly, ten to twenty picoseconds after the end of the exciting laser pulse. This value of $\tau_3$ is expected to be quite similar for most direct-gap polar semiconductors, but may differ in other types of HEPS, each case requiring a calculation like the one described here.

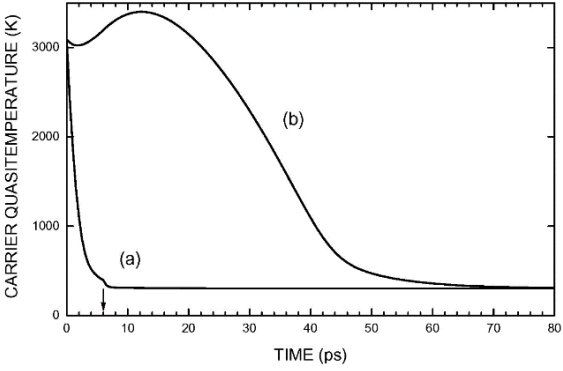

**Figure 8.** Comparison of the evolution of the carriers' temperature when (a) is taken a unique quasitemperature for the phonons modes, and (b) the one when it is calculated the evolution of the quasitemperature of each phonon mode. As shown in Figure 10 the latter agrees quite well with the experimental data. (Algarte[70]).

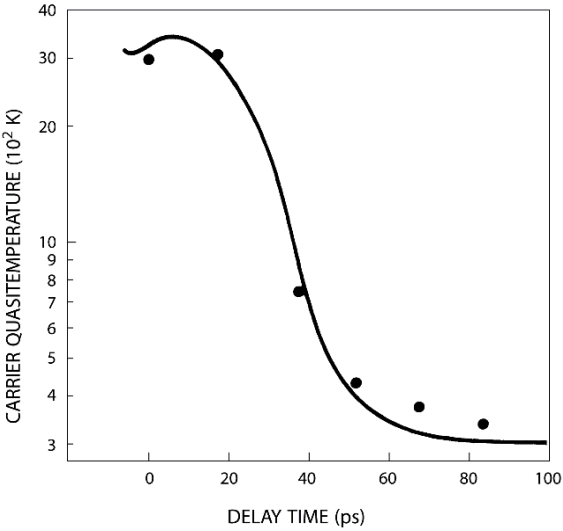

**Figure 9.** Evolution of the carrier quasitemperature in highly excited GaAs. The NESEF-based calculation (full line) is from Daly & Mahr[81] and experimental data (dots) are from Seymour et al.[72].

The experiments of Leheny et al.[73], von der Linde & Lambrich[74] and Seymour et al.[72] provide data that were analyzed in terms of a description of the HEPS in the third kinetic stage (ten or more picoseconds after the end of the exciting laser pulse). Figures 10a, 10b and 10c show a comparison of theoretical (full-lines) and experimental (dots) results displaying very good agreement[75]. This indicates that the contracted description of the third kinetic stage applies well and thermalization of carriers and optical phonons has been established. In Figure 10c the dashed curve is a calculation neglecting the anharmonic interaction among phonons. The much better agreement shown by the full curve (including anharmonic interactions) is a clear manifestation of our previous arguments that the transfer of the carrier excess energy to the reservoir is channeled via optical phonons and anharmonic interaction.

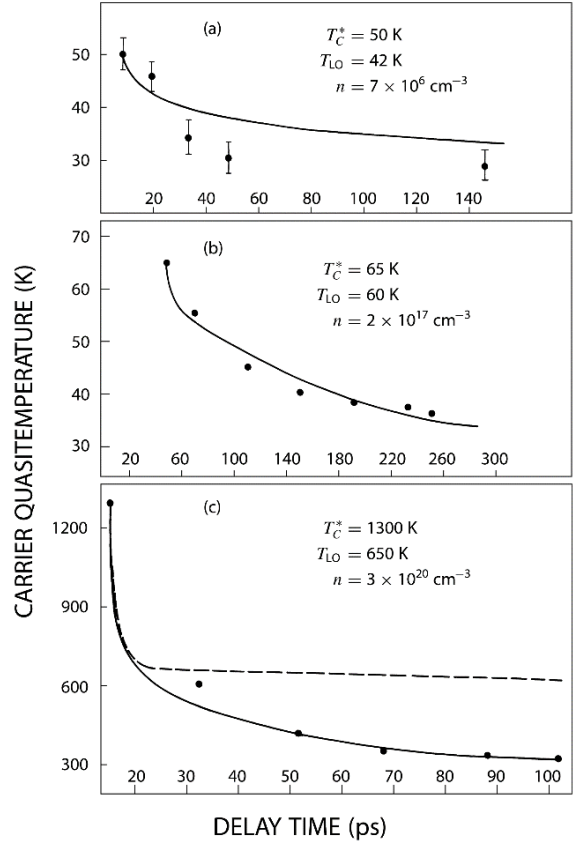

**Figure 10.** Evolution of the carriers' quasitemperature in three different experimental situations (indicated in the main text) which are the full lines calculated in NESEF, compared with experimental data (dots) (Sampaio[75]).

We recall that the carrier quasitemperature is derived from the slope of the high-frequency side of the luminescence spectrum, $I(\omega|t)$[76]. It follows that

$$\left| \frac{d}{d\omega} \ln I(\omega|t) \right| = \frac{\hbar}{k_B T_c^*(t)} \qquad (6)$$

(in modulus because the slope is negative) providing what can be called a "thermometric device" for the measurement of the carriers' quasitemperature.

## 2.5. Relaxation towards final thermodynamic equilibrium

After the third kinetic stage has been set, the coincident quasitemperatures $T_c^*(t) = T_{LO}^*(t) = T_{TO}^*(t)$ keep decreasing at a slow pace towards final thermal equilibrium at the reservoir temperature $T_0$. However, it must be kept in mind that electron-hole recombination is present with a recombination time of the order of nanoseconds. Hence, during relaxation to final equilibrium the concentration of carriers will decrease enough to lead to a Mott-phase transition from the HEPS metallic phase to the excitonic non-metallic phase, with eventual formation of excitonic complexes. At very low temperatures, a Keldysh phase of an exciton gas in coexistence with metallic electron-hole droplets[77] may even occur. In Figure 9 is evident the evolution of the carriers' quasitemperature towards a final thermal equilibrium with the reservoirs at 300 K, following in a larger than 100 ps scale after application of the laser pulse. The full line is the calculation in NESEF and the dots are experimental data taken from Seymour et al.[72]. We recall that the carriers' quasitemperature is measured using the value of the angular coefficient, in a logarithmic plot, of the high frequency side of the luminescent spectra (cf. Equation 6)[78].

Similar situation can be observed in Figure 11 in the case of the experiment of Amand & Collet[79]; again the full line is the calculation in NESEF and the dots are experimental data. The dashed curve is a calculation when ambipolar diffusion of electrons and holes out of the active volume of the sample (the region of focusing of the laser beam) is neglected, resulting in a more rapid relaxation, contrary to what is observed; the arrow indicates the end of the exciting laser pulses, which was taken with a rectangular profile. This then evidences the influence in the change of the density in the photoinjected plasma, a change illustrated in Figure 12, (showing the calculation and the experimental data), resulting from radiative recombination and, mainly, the aforementioned ambipolar diffusion[80]. It can be noticed that the latter is also responsible for the emergence of the shoulder that can be seen in the plot in Figure 10.

Figure 13 provides a summarized picture of the contraction processes in the macroscopic description of the homogeneous HEPS here described. We have illustrated the matter considering experiments involving ultrafast-time resolved luminescence. In Figure 13 it is shown a set of spectra taken at increasing times (resolution is in the order of some nanoseconds) of the luminescence in platelets of CdS[81]. This is quite illustrative of the evolution of fast irreversible processes, what is evidenced in the observation that, (1) the width of the band decreases in time, while (2) the slope of the spectra in the high frequency side increases, and (3) there is a shift in the position of the peak. This is a result that as the nonequilibrium state of the system evolves in time, the concentration of electron-hole pairs decreases (tending to return to the equilibrium value) and the energy in excess of equilibrium is also returning – via relaxation processes to the lattice – to its value in equilibrium[82,83].

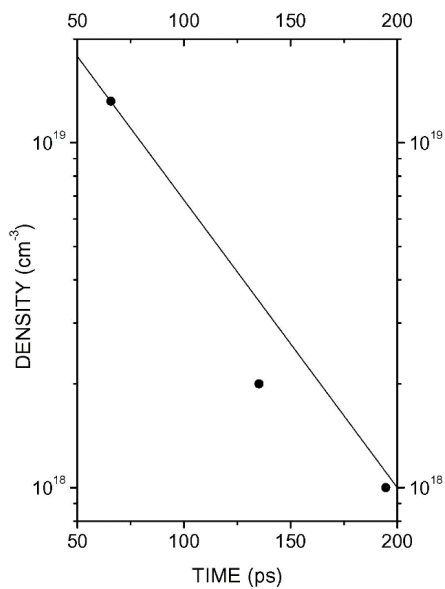

**Figure 12.** Evolution of the photoinjected carrier concentration in the same conditions of Figure 11, governed mainly by ambipolar diffusion. Full line is the NESEF-based calculation and the dots are experimental data from Amand & Collet[79].

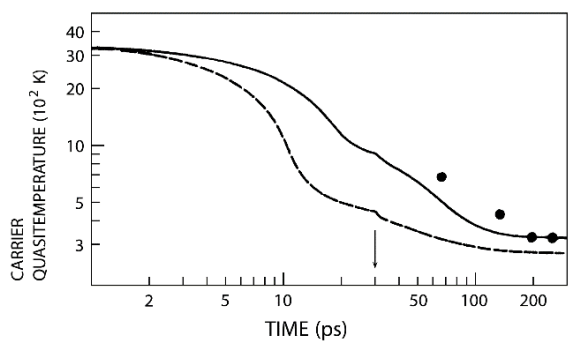

**Figure 11.** Evolution of the carrier quasitemperature in highly excited GaAs. The NESEF-based calculation (full line) is from Algarte et al.[82] and experimental data (dots) are from Amand & Collet[79].

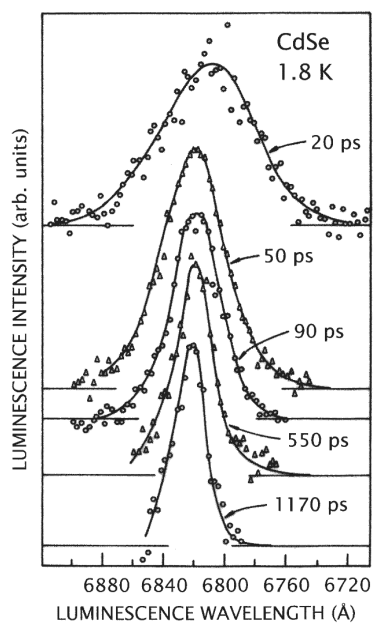

**Figure 13.** Photoluminescence spectra of CdSe platelets under high levels of excitation, obtained at the indicated delay times after pulse excitation. (Daly[81]).

# 3. Time Resolved Optical Properties in HEPS

It may be noticed that for the study of the ultrafast time evolving optical properties in the plasma in semiconductors it is necessary to derive in detail a response function theory and scattering theory for systems far from equilibrium.

Particularly, one needs to derive the frequency- and wave number-dependent dielectric function in arbitrary nonequilibrium conditions, because it is the quantity which contains all the information related to the optical properties of the system (as known, it provides the absorption coefficient, the reflectivity coefficient, the Raman scattering cross section, etc.). We describe the application of the results to the study of a particular type of experiment, namely the time-resolved reflectivity changes in GaAs and other materials in Albrecht et al.[84], Cho et al.[85], Pfeifer et al.[86] and Scholz et al.[87] where signal changes in the reflectivity, $\Delta R/R$, of the order of $10^{-7}$ are detected, and a distinct oscillation of the signal in real time is observed. In Figure 14 are reproduced time-resolved reflectivity spectra, and in the upper right inset is shown the part corresponding to the observed oscillation, as reported by Cho et al.[85].

Such phenomenon has been attributed to the generation of coherent lattice vibrations, and several theoretical approaches have been reported[87-90]. A clear description, on phenomenological bases, which captures the essential physics of the problem, is reported in Zeiger et al.[90], and in Vasconcellos et al.[91] is presented an analysis based on NESEF, where it is evidenced that the oscillatory effect is provided by the displacive excitation of the polar lattice vibrations, arising out of the coupling of the carrier-charge density and polar modes, and its decay is mainly governed by the cooling down of the carriers, and where the different physical aspects of the problem are discussed. We briefly describe next the full use of NESEF for dealing with pump-probe experiments for studying the optical properties of semiconductors in nonequilibrium conditions.

Let us consider a direct-gap polar semiconductor in a pump-probe experiment. We recall that the exciting intense laser pulse produces the so-called highly excited plasma in semiconductors HEPS (see Figure 2), namely, electron-hole pairs in the metallic side of Mott transition (that is, they are itinerant carriers, and we recall that this requires concentrations of these photoinjected quasi-particles of order of $10^{16}$cm$^{-3}$ and up), which compose a two-component Fermi fluid, moving in the lattice background. It constitutes a highly nonequilibrated system where the photoexcited carriers rapidly redistribute their energy in excess of equilibrium via, mainly, the strong long-range Coulomb interaction (pico- to subpico- second scale), followed by the transfer of energy to the phonon field (predominantly to the optical phonons, and preferentially to the LO phonons via Fröhlich interaction), and finally via acoustic phonons to the external thermal reservoir, as we have already described. Along the process the carrier density diminishes in recombination processes (nanosecond time scale) and through ambipolar diffusion out of the active volume of the sample (ten-fold picosecond time scale).

Moreover, a probe interacting weakly with the HEPS is used to obtain an optical response, the reflectivity of the incoming laser photons with frequency $\omega$ and wavevector $\mathbf{Q}$ in the case under consideration. From the theoretical point of view, such measurement is to be analyzed in terms of the all important and inevitable use of correlation functions in response function theory[52,53,92]. The usual application in normal probe experiments performed on a system initially in equilibrium had a long history of success, and a practical and elegant treatment is based on the method of the double-time (equilibrium) thermodynamic Green functions[52,53]. In the present case of a pump-probe experiment we need to resort to a theory of such type but applied to a system whose macroscopic state is in nonequilibrium conditions and evolving in time as a result of the pumping dissipative processes that are developing while the sample is probed, that is, the response function theory for nonequilibrium systems, which needs be coupled to the kinetic theory that describes the evolution of the nonequilibrium state of the system[18,19,93,94]. We resort here to such theory for the study of the optical properties in HEPS, and, in particular, we consider the case of reflectivity.

The time-dependent (because it keeps changing along with the evolution of the macrostate of the nonequilibrated system) $R(\omega,\mathbf{Q}|t)$ is related to the index of refraction $\eta(\omega,\mathbf{Q}|t) + i\kappa(\omega,\mathbf{Q}|t)$ through the well-known expression

$$R(\omega,\mathbf{Q}|t) = \frac{[\eta(\omega,\mathbf{Q}|t)-1]^2 + [\kappa(\omega,\mathbf{Q}|t)]^2}{[\eta(\omega,\mathbf{Q}|t)+1]^2 + [\kappa(\omega,\mathbf{Q}|t)]^2} \quad (7)$$

and the refraction index is related to the time-evolving frequency- and wavevector-dependent dielectric function by

$$\epsilon(\omega,\mathbf{Q}|t) = \epsilon'(\omega,\mathbf{Q}|t) + i\epsilon''(\omega,\mathbf{Q}|t) = [\eta(\omega,\mathbf{Q}|t) + i\kappa(\omega,\mathbf{Q}|t)]^2 \quad (8)$$

where $\eta$ and $\epsilon'$, and $\kappa$ and $\epsilon''$, are the real and imaginary parts of the refraction index and of the dielectric function.

We call the attention to the fact that the dielectric function depends on the frequency and the wavevector of the radiation involved, and $t$ stands for the time when a measurement is performed. Once again we stress that this dependence on time is, of course, the result that the macroscopic state of the non-equilibrated plasma is evolving in time as the experiment is performed.

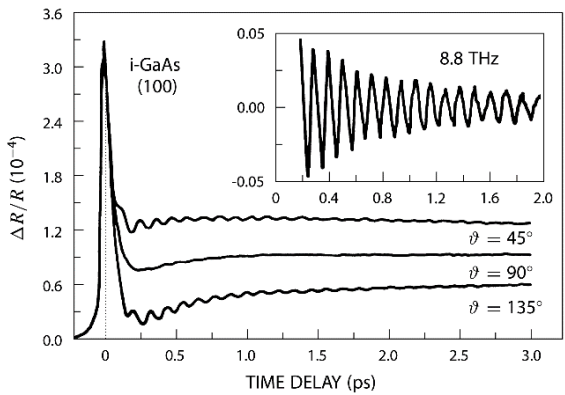

**Figure 14.** Reproduction of the time-resolved reflectivity changes in GaAs, as reported by Cho et al.[85].

Therefore it is our task to calculate this dielectric function in the nonequilibrium state of the HEPS. First, we note that according to Maxwell equations in material media (that is, Maxwell equations now averaged over the nonequilibrium statistical ensemble) we have that

$$\epsilon^{-1}(\omega, \mathbf{Q} | t) - 1 = \frac{n(\omega, \mathbf{Q} | t)}{r(\omega, \mathbf{Q})} \quad (9)$$

where $r(\omega, \mathbf{Q})$ is the amplitude of a probe charge density with frequency $\omega$ and wavevector $\mathbf{Q}$, and $n(\omega, \mathbf{Q}|t)$ the induced polarization-charge density of carriers and lattice in the media. The latter can be calculated resorting to the response function theory for systems far from equilibrium (the case is quite similar to the calculation of the time-resolved Raman scattering cross section[95], and obtained in terms of the nonequilibrium-thermodynamic Green functions, as we proceed to describe.

Using the formalism described in Vasconcellos et al.[91] to obtain $\epsilon(\omega, \mathbf{Q}|t)$, it follows that,

$$\epsilon^{-1}(\omega, \mathbf{Q}) - 1 = V(\mathbf{Q})[G_{cc}(\omega, \mathbf{Q}) + G_{ci}(\omega, \mathbf{Q}) + G_{ic}(\omega, \mathbf{Q}) + G_{ii}(\omega, \mathbf{Q})] \quad (10)$$

giving the reciprocal of the dielectric function in terms of Green functions given by

$$G_{cc}(\omega, \mathbf{Q}) = \langle\langle \hat{n}_c(\mathbf{Q}); \hat{n}_c^\dagger(\mathbf{Q}) | \omega; t \rangle\rangle \quad (11)$$

$$G_{ci}(\omega, \mathbf{Q}) = \langle\langle \hat{n}_c(\mathbf{Q}); \hat{n}_i^\dagger(\mathbf{Q}) | \omega; t \rangle\rangle \quad (12)$$

$$G_{ic}(\omega, \mathbf{Q}) = \langle\langle \hat{n}_i(\mathbf{Q}); \hat{n}_c^\dagger(\mathbf{Q}) | \omega; t \rangle\rangle \quad (13)$$

$$G_{ii}(\omega, \mathbf{Q}) = \langle\langle \hat{n}_i(\mathbf{Q}); \hat{n}_i^\dagger(\mathbf{Q}) | \omega; t \rangle\rangle \quad (14)$$

where $V(\mathbf{Q}) = 4\pi n e^2 / V \epsilon_0 Q^2$ is the matrix element of the Coulomb potential in plane-wave states and $\hat{n}_c(\mathbf{Q})$, and $\hat{n}_i(\mathbf{Q})$, refer to the $\mathbf{Q}$-wave vector Fourier transform of the operators for the densities of charge of carriers and ions respectively.

But, the expression we obtain is, as already noticed, depending on the evolving nonequilibrium macroscopic state of the system, a fact embedded in the expressions for the time-dependent distribution functions of the carrier and phonon states. Therefore, they are to be derived within the kinetic theory in NESEF, and the first and fundamental step is the choice of the set of variables deemed appropriate for the description of the macroscopic state of the system. The case of HEPS has already been discussed, and we simply recall that a first set of variables needs be the one composed of the carriers' density and energy, and the phonon population functions, together with the set of associated nonequilibrium thermodynamic variables that, as we have seen, can be interpreted as a reciprocal quasitemperature and quasi-chemical potentials of carriers, and reciprocal quasitemperatures of phonons, one for each mode. But in the situation we are considering we need to add, on the basis of the information provided by the experiment, the amplitudes of the LO-lattice vibrations and the carrier charge density; the former because it is clearly present in the experimental data (the oscillation in the reflectivity) and the latter because of the LO-phonon-plasma coupling (clearly present in Raman scattering experiments[95-99]). Consequently the chosen basic set of dynamical quantities is

$$\{\hat{H}_c, \hat{N}_e, \hat{N}_h, \hat{n}_{\mathbf{kp}}^e, \hat{n}_{\mathbf{kp}}^h, \hat{v}_{\mathbf{q}}, a_{\mathbf{q}}, a_{\mathbf{q}}^\dagger, H_B\} \quad (15)$$

where

$$\hat{H}_c = \sum_{\mathbf{k}} \left[ \varepsilon_{\mathbf{k}}^e c_{\mathbf{k}}^\dagger c_{\mathbf{k}} + \varepsilon_{\mathbf{k}}^h h_{-\mathbf{k}}^\dagger h_{-\mathbf{k}} \right] \quad (16)$$

$$\hat{v}_{\mathbf{q}} = a_{\mathbf{q}}^\dagger a_{\mathbf{q}} \quad (17)$$

$$\hat{N}_e = \sum_{\mathbf{k}} c_{\mathbf{k}}^\dagger c_{\mathbf{k}}, \quad \hat{n} N_h = \sum_{\mathbf{k}} h_{-\mathbf{k}}^\dagger h_{-\mathbf{k}} \quad (18)$$

$$\hat{n}_{\mathbf{kp}}^e = c_{\mathbf{k}+\mathbf{p}}^\dagger c_{\mathbf{k}}, \quad \hat{n}_{\mathbf{kp}}^h = h_{-\mathbf{k}-\mathbf{p}} h_{-\mathbf{k}}^\dagger \quad (19)$$

with $c$ ($c^\dagger$), $h$ ($h^\dagger$), and $a$ ($a^\dagger$) being as usual annihilation (creation) operators in electron, hole, and LO-phonon states respectively (**k**, **p**, **q** run over the Brillouin zone). Moreover, the effective mass approximation is used and Coulomb interaction is dealt with in the random phase approximation, and then $\epsilon_{\mathbf{k}}^e = E_G + \hbar^2 |\mathbf{k}|^2 / 2m_e^*$ and $\epsilon_{\mathbf{k}}^h = \hbar^2 |\mathbf{k}|^2 / 2m_h^*$. Finally $H_B$ is the Hamiltonian of the lattice vibrations different from the LO one. We write for the NESEF-nonequilibrium thermodynamic variables associated to the quantities of Equation 15

$$\{\beta_c(t), -\beta_c^*(t)\mu_e^*(t), -\beta_c^*(t)\mu_h^*(t), F_{\mathbf{kp}}^e(t), \\ F_{\mathbf{kp}}^h(t), \hbar\omega_{\mathbf{q}}\beta_{\mathbf{q}}^*(t), \varphi_{\mathbf{q}}(t), \varphi_{\mathbf{q}}^*(t), \beta_0\} \quad (20)$$

respectively, where $\mu_e^*$ and $\mu_h^*$ are the quasi-chemical potentials for electrons and for holes; we write $\beta_c^*(t) = 1/k_B T_c^*(t)$ introducing the carriers' quasitemperature $T_c^*$; $\beta_{\mathbf{q}}^*(t) = 1/k_B T_{\mathbf{q}}^*(t)$ introducing the LO-phonon quasitemperature per mode ($\omega_{\mathbf{q}}$ is the dispersion relation)[54,72,78], $\beta_0 = 1/k_B T_0$ with $T_0$ being the temperature of the thermal reservoir. We indicate the corresponding macrovariables, that is, those which define the nonequilibrium thermodynamic space as

$$\{E_c(t), n(t), n(t), n_{\mathbf{kp}}^e(t), n_{\mathbf{kp}}^h(t), v_{\mathbf{q}}(t), \\ \langle a_{\mathbf{q}}|t\rangle, \langle a_{\mathbf{q}}^\dagger|t\rangle = \langle a_{\mathbf{q}}|t\rangle^*, E_B\} \quad (21)$$

which are the statistical average of the quantities of Equation 15, that is

$$E_c(t) = \text{Tr}\{\hat{H}_c \varrho_\epsilon(t) \times \varrho_R\} \quad (22)$$

$$n(t) = \text{Tr}\{\hat{N}_{e(h)} \varrho_\epsilon(t) \times \varrho_R\} \quad (23)$$

and so on, where $\varrho_R$ is the stationary statistical distribution of the reservoir and $n(t)$ is the carrier density, which is equal for electrons and for holes since they are produced in pairs in the intrinsic semiconductor. The volume of the active region of the sample (where the laser beam is focused) is taken equal to 1 for simplicity.

Next step is to derive the equation of evolution for the basic variables that characterize the nonequilibrium macroscopic state of the system, and from them the evolution of the nonequilibrium thermodynamic variables. This is done according to the NESEF generalized nonlinear quantum transport theory, but in the second-order approximation in relaxation theory. This is an approximation which retains only two-body collisions but with memory and vertex renormalization being neglected, consisting in the Markovian limit of the theory. It is sometimes referred to as the quasi-linear approximation in relaxation theory[100], a name we avoid because of the misleading word linear which refers to the lowest order in dissipation, however the equations are highly nonlinear.

The NESEF auxiliary ("instantaneously frozen") statistical operator is in the present case given, in terms of the variables of Equation 15 and nonequilibrium thermodynamic variables of Equation 20, by

$$\bar{\varrho}(t,0) = \exp\{-\phi(t) - \beta_c(t)[\hat{H}_c - \mu_e(t)\hat{N}_e - \mu_h(t)\hat{N}_h] - \sum_{\mathbf{kp}}[F^e_{\mathbf{kp}}(t)\hat{n}^e_{\mathbf{kp}}(t) + F^h_{\mathbf{kp}}(t)\hat{n}^h_{\mathbf{kp}}(t)] - \sum_{\mathbf{kp}}[\beta_{\mathbf{q}}(t)\hbar\omega_{\mathbf{q}}\hat{v}_{\mathbf{q}} + \varphi_{\mathbf{q}}(t)a_{\mathbf{q}} + \varphi^*_{\mathbf{q}}(t)a^\dagger_{\mathbf{q}}] - \beta_0 H_B\} \quad (24)$$

where $\phi(t)$ ensures the normalization of $\bar{\varrho}(t,0)$. Moreover, we recall, the statistical operator $\varrho_\varepsilon(t)$ is given by

$$\varrho_\varepsilon(t) = \bar{\varrho}(t,0) + \bar{h}'_\varepsilon(t) = \exp\{-\hat{S}(t,0) + \int_{-\infty}^{t} dt' e^{\varepsilon(t'-t)} \frac{d}{dt}\hat{S}(t',t'-t)\} \quad (25)$$

where

$$\hat{S}(t,0) = -\ln\bar{\varrho}(t,0) \quad (26)$$

$$\hat{S}(t',t'-t) = \exp\left\{-\frac{1}{i\hbar}(t'-t)\hat{H}\right\}\hat{S}(t',0)\exp\left\{-\frac{1}{i\hbar}(t'-t)\hat{H}\right\} \quad (27)$$

with $\hat{S}(t,0)$ being the so-called informational-statistical entropy operator, and $\hat{H}$ stands as usual for the system Hamiltonian (which includes the interaction with the external sources and with the thermal reservoir). Moreover, and for the sake of completeness, we recall that in Equation 25 $\varepsilon$ is a positive infinitesimal that goes to zero after the calculation of averages: This procedure introduces Bogoliubov's quasiaverage method which in this case breaks the time-reversal symmetry of Liouville equation introducing irreversible behavior from the onset[18,19,101-106].

Using the statistical operator of Equation 25 the Green functions that define the dielectric function [cf. Equation 10 can be calculated. This is an arduous task, and in the process it is necessary to evaluate the occupation functions

$$f_\mathbf{k}(t) = \text{Tr}\{c^\dagger_\mathbf{k} c_\mathbf{k} \varrho_\varepsilon(t)\} \quad (28)$$

which for weak inhomogeneities, as it is the present case, is presented in Vasconcellos et al.[91]. The (nonequilibrium) carrier quasitemperature $T^*_c$ is obtained, and its evolution in time shown in Figure 15.

Finally, in Figure 16, leaving only as an adjustable parameter the amplitude – which is fixed fitting the first maximum –, is shown the calculated modulation effect which is compared with the experimental result (we have only placed the positions of maximum and minimum amplitude taken from the experimental data, which are indicated by the full squares).

In that way this demonstrates the reason of the presence of the observed modulating phenomenon in the reflectivity spectra, occurring with the frequency of the near zone center LO-phonon (more precisely the one of the upper $L_+$ hybrid mode[96] with wave vector **Q**, the one of the photon in the laser radiation field. The amplitude of the modulation is determined by the amplitude of the laser-radiation-driven carrier charge density which is coupled to the optical vibration, and then an open parameter in the theory to be fixed by the experimental observation. This study has provided, as shown, a good illustration of the full use of the formalism of NESEF, with an application to a quite interesting experiment and where, we recall, the observed signal associated to the modulation is seven orders of magnitude smaller than the main signal on which is superimposed.

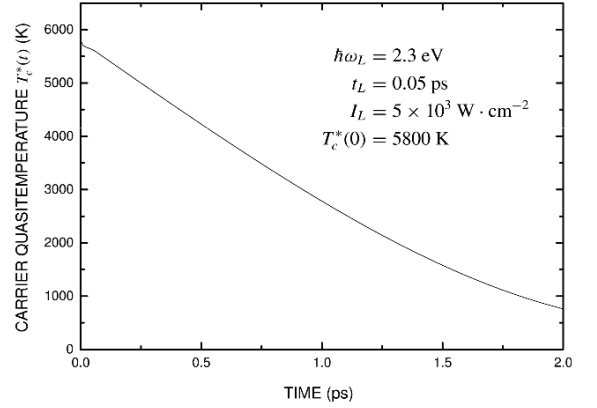

**Figure 15.** Evolution of the carrier's quasitemperature, calculated in the conditions of the experiment in the caption to Figure 14 (Vasconcellos[91]).

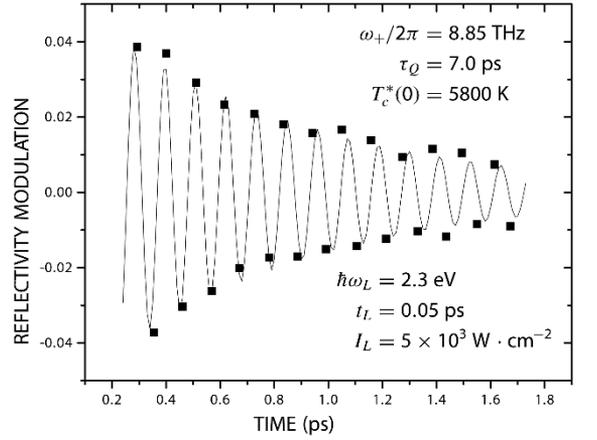

**Figure 16.** The theoretically evaluated modulation of the time-resolved reflectivity in the conditions of Cho et al.[85], compared with the experimental data. For simplicity we have drawn only the positions of the maxima and minima of the figure in the inset of Figure 14 (Vasconcellos[91]).

## 4. Concluding Remarks

As already noticed, nowadays advanced electronic and opto-electronic semiconductor devices work in far-from-equilibrium conditions, involving ultrafast relaxation processes (in the pico- and femto-second scales) and ultrashort (nanometric) scales (constrained geometries). We have presented here a detailed analysis of the physics involved in the evolution of ultrafast relaxation processes in polar semiconductors. This was done in the framework of the Non-Equilibrium Statistical Ensemble Formalism which allows for the proper description of the ultrafast evolution of the macroscopic non-equilibrium thermodynamic state of the system.

The theory was applied to the analysis of several experiments in the field of ultrafast laser spectroscopy. In a future article we shall present the use of NESEF for dealing with transport properties of polar semiconductors in the presence of moderate to high electric fields, and also the possibility of emergence of complex behavior.

## Acknowledgements

The authors would like to acknowledge partial financial support received from the São Paulo State Research Agency (FAPESP) and the Brazilian National Research Council (CNPq): The authors are CNPq Research Fellows. The author CGR thanks the financial support received from Goiás State Research Agency (FAPEG).